\documentclass[aps,prd,nofootinbib,floatfix,twocolumn,showpacs]{revtex4-1}
\usepackage{graphicx,amsmath,amssymb}
\usepackage{psfrag}
\usepackage{color}
\usepackage{hyperref}
\newcommand{\be}{\begin {eqnarray}}
\newcommand{\ee}{\end{eqnarray}}
\newcommand{\up}{\uparrow}
\newcommand{\down}{\downarrow}
\newcommand{\ketep}{\vert\,{e_\text{phys}}\rangle}

\newcommand{\braket}[2]{\bra{#1}\,#2\rangle} 
\newcommand{\bra}[1]{\langle\,#1\,|}          
\newcommand{\ket}[1]{|\,#1\,\rangle}          
\newcommand{\ud}{\mathrm{d}}
\newcommand{\ie}{\textit{i.e.}} 

\newcommand{\LCm}{{\scriptscriptstyle -}} 
\newcommand{\LCp}{{\scriptscriptstyle +}}

\newcommand{\LCperp}{{\scriptscriptstyle \perp}}

\begin{document}
\title{Generalized parton distributions in a light-front  nonperturbative approach }
\author{\bf D. Chakrabarti$^1$, X.  Zhao$^2$, H. Honkanen$^3$ R. Manohar$^4$, P. Maris$^2$,  J. P. Vary$^2$ }
\affiliation{$^1$Dept. of Physics,  Indian Institute of Technology Kanpur, Kanpur-208016, India \\
$^2$Dept. of Physics and Astronomy, Iowa State University, Ames, IA 50011, USA\\
$^3$Dept. of Physics, Penn State University, University Park, PA 16802, USA\\
$^4$Dept. of Physics, BITS-Pilani (Goa Campus), Goa 403726, India.}
\date{\today}
\begin{abstract}
Basis Light-front Quantization (BLFQ) has recently been developed as a promising nonperturbative technique.  Using BLFQ, we investigate the Generalized Parton Distributions (GPDs) in a nonperturbative framework for a dressed electron in QED. We evaluate light-front wave functions and carry out overlap calculations to obtain GPDs.  We also perform perturbative calculations in the corresponding basis spaces to demonstrate that they compare reasonably with the BLFQ results.
 \end{abstract}


\maketitle

\section{Introduction}

More than forty years after  the advent of QCD, the understanding of  hadronic structures
is incomplete.  Since the subject involves strong interactions, we are handicapped by the lack of nonperturbative techniques to investigate these structures  in a practical manner.  Lattice gauge theory provides us  the most practiced nonperturbative method to address the issues in QCD, but it has limitations. Recently Basis Light-front Quantization (BLFQ) \cite{BLFQ1,BLFQ2} has been developed as another nonperturbative tool and has been extended and applied to investigate  scattering \cite{Zhao:2013cma,Zhao:2013jia} and bound state problems \cite{BLFQ3}.

  In this work, we calculate the generalized parton distributions (GPDs) which encode  nonperturbative information about  the spatial structure of the hadron as well as the spin and angular momentum contributions of the partons to the hadron. GPDs appear in the amplitudes for exclusive processes like Deeply Virtual Compton Scattering (DVCS) or Deeply Virtual vector Meson Production (DVMP).  There are several extensive reviews and discussions about GPDs in the literature \cite{MD,BR,GPV,BP}.
   The GPDs are functions of the longitudinal momentum fraction of the probed parton ($x$), the longitudinal momentum fraction transferred, or skewness, ($\zeta$) and the square of the momentum transferred ($t$).  In the forward limit ($t\to 0$) they reduce to the ordinary parton distribution functions (PDFs). As they are off-forward matrix elements, the GPDs are not positive definite and therefore cannot be interpreted as  ``distribution functions".  However, in the zero skewness limit, the Fourier transform  of the GPDs with respect to the transverse momentum transfer ($\Delta_\perp$) gives the GPDs in  impact parameter space which are positive definite and provides the distribution of the partons in transverse position space \cite{burk}. 
Ji's sum rule \cite{ji} relates the  second moment of the sum of GPDs $H(x,\zeta,t)$ and $E(x,\zeta,t)$ in the $t\to 0$ limit with the total angular momentum of the parton.  For a transversely polarized proton, one can interpret  the term $H(x,0,0)$ in Ji's sum rule  as the effect of  an overall transverse shift when going from instant form to front form whereas  the term $E(x,0,0)$ arises from the transverse deformation of the GPDs in the center of momentum frame~\cite{burk2}.  There are many existing experimental results \cite{expts}  and  also  many future experiments  that will study  DVCS/DVMP. These experiments require a model to extract the  GPDs for the proton   which has prompted  many theoretical  and phenomenological models for GPDs \cite{models1,models2}. The nonperturbative  lattice approach can  calculate  a limited set of   moments of GPDs \cite{lattice} and therefore requires additional assumptions for higher moments to generate the complete GPDs.

In this work, we report an initial effort to investigate  the GPDs  in a completely nonperturbative framework.  In particular,  we  calculate the GPDs of an electron dressed with a photon at zero skewness. Since we work initially in QED, we exploit the opportunity to compare our nonperturbative results with perturbative results.  The  light-front framework allows us  to expand the physical state into a multi-particle  Fock space. The GPDs in this QED problem satisfy the nontrivial  properties like polynomiality  (for nonzero $\zeta$) and positivity conditions which are reasons why the dressed electron in  QED has  been  extensively used to study DVCS and  GPDs~\cite{ BDH, hadron, CM,CMM,MV}.

This paper is organized as follows.  In Sec.\ref{overlap}, we briefly outline  the  light-front overlap representation  of the GPDs and in Sec.\ref{blfq} we introduce the basics of BLFQ and the procedure of evaluating electron GPDs in BLFQ. Then in Sec.\ref{model}, we present our model light-front wave functions and the resulting GPDs in perturbation theory. In Sec.\ref{results}, we compare our numerical results from BLFQ to those from perturbative calculations. Finally we finish the paper with a summary and conclusions. 


\section{Overlap Representation\label{overlap}}

In this work, as a test problem for BLFQ, instead of considering the GPDs of a hadron in QCD we focus on the GPDs of a physical (dressed) electron in QED  and 
the corresponding ``partons'' are the bare electrons and photons.
We expand the physical  electron into the  Fock space basis:
\begin{widetext}
\begin{align}
\label{eq:phys_e_Fock_expansion}
\ketep=&\psi_1\mid e^-\rangle+\psi_2\mid e^- \gamma\rangle+\psi_{3a}\mid e^-e^+e^-\rangle+\psi_{3b}\mid e^-\gamma\gamma\rangle+\cdots
\end{align}
We choose the frame where the initial and final momenta of the electron with  mass $M$ are:
\be
P&=&\left( P^+, 0_\perp, {M^2\over P^+ }\right),\\
P'&=&\left( (1-\zeta)P^+, -\Delta_\perp, {M^2+\Delta_\perp^2\over 
(1-\zeta)P^+ } \right).
\ee
So, the momentum transferred from the target is
\be
\Delta=P-P'=\left( \zeta P^+, \Delta_\perp, {t+\Delta_\perp^2 \over \zeta P^+} 
\right),
\ee
where $t=\Delta^2$.

The GPDs are expressed as the off-forward matrix element of the
bilocal  operator on the light cone. 
We use the parameterization   for the  GPDs~\cite{BDH}:
\be
  &&\int {dz^-\over 8 \pi} e^{i xP^+z^-/2} \langle P' \mid
{\bar \Psi} ( {0})\gamma^{+}  \Psi ({z^- }) \mid P \rangle\mid_{z^+=0,z^\perp=0}
\nonumber\\
&& ~~~~= \frac{1}{2 {\bar P}^+}\Big(H^q(x,\zeta,t) {\bar u} (P') \gamma^{+}  u(P) 
+  E^q (x,\zeta,t) {\bar u}(P') \frac{i\sigma^{+j}(-{\Delta_j})}{2 M} 
u(P) \Big), \label{GPD}
\ee       
where ${\bar P}^+=(P^++P'^+)/2=(1-\zeta/2)P^+$. The physical electron state of momentum  $P$ is then expanded in terms of
multi-particle light-front wave functions \cite{BDH}:
\be
\mid P \rangle &=& \sum_{n} \int \prod_{i=1}^n {dx_i d^2 k_{\perp i} \over
\sqrt{x_i} 16 \pi^3} 16 \pi^3 \delta\left(1- \sum_{i=1}^n x_i\right) \delta^2
\left(\sum_{i=1}^n k_{\perp i}\right)\psi_n(x_i, k_{\perp i}, \lambda_i)
\mid n, x_i P^+, x_i P_\perp + k_{\perp i}, \lambda_i \rangle;
\ee
here $x_i=k_i^+/P^+$   and $k_{\perp i}$ 
represent the relative transverse momentum of the $i$-th constituent and $n$ is the number of particles in a Fock state. The
physical transverse momenta are $p_{\perp i} = x_i P_\perp + k_{\perp i}$ and
$\lambda_i$ are the light-cone helicities. The boost invariant light-front wave functions 
$\psi_n$ depend only on $x_i$ and  $k_{\perp i}$  and are independent of the total momentum of the state $P^+$ and $P_\perp$.     

In this work, we concentrate only on the zero skewness limit, \ie, $\zeta=0$.  In this limit,  for $0\le x\le 1$, only the diagonal (\ie, $n\to n$) process contributes to the GPDs and the overlap representations of the GPDs are given by:
\be
H^q(x,0,t)&=& 
 \sum_{n,\lambda_i} \int \prod_{i=1}^n {dx_i d^2
k_{\perp i}\over 16 \pi^3} 16 \pi^3 \delta\left(1-\sum_j x_j\right)
\delta^2\left(\sum_{j=1}^n k_{\perp j}\right) \delta(x-x_1) \nonumber\\&& \psi_n^{\uparrow*}(x_i',
{k'}_{\perp i},\lambda_i) \psi_n^{\uparrow}(x_i,
k_{\perp i},\lambda_i) ;    \\
\frac{\Delta^1-i\Delta^2}{2 M}E^q(x,0,t)&=& 
 \sum_{n,\lambda_i} \int \prod_{i=1}^n {dx_i d^2
k_{\perp i}\over 16 \pi^3} 16 \pi^3 \delta\left(1-\sum_j x_j\right)
\delta^2\left(\sum_{j=1}^n k_{\perp j}\right) \delta(x-x_1) \nonumber\\&& \psi_n^{\uparrow*}(x_i',
{k'}_{\perp i},\lambda_i) \psi_n^{\downarrow}(x_i,
k_{\perp i},\lambda_i) ; 
\ee
\end{widetext}
where  for the struck parton 
 ${x'}_1=x_1; 
~{k'}_{\perp 1}=k_{\perp 1}-(1-x_1) \Delta_\perp$ and  ${x'}_i={x_i}; ~{k'}_{\perp i}=k_{\perp i}+{x_i} \Delta_\perp$ for  the spectators ($i=2,....n$).

\section{Basis Light Front Quantization\label{blfq}}
In this section we present a brief outline of the Basis Light-front Quantization (BLFQ) (see Refs.~\cite{BLFQ1,Zhao:2013cma} for more details). BLFQ is a first-principles nonperturbative approach to bound state problems in quantum field theory (QFT). BLFQ adopts the light-front quantization~\cite{DLCQ},  and 
the bound state problem in QFT is treated as the eigenvalue problem of the light-front Hamiltonian of the system. Explicitly, through solving the following eigenequation of the light-front Hamiltonian, $P^-$,
\begin{align}
P^-\ket{\beta}=P^-_\beta\ket{\beta},
\end{align}
one obtains the (light-front) energy spectrum, $P^-_\beta$, and the associated (light-front) amplitudes 
of the bound states. The (squared) invariant mass of bound states, $M^2$, is related to the light-front energy according to,
\begin{align}
M^2 = P^+P^--P_\perp^2,
\end{align}
where $P_\perp$ ($P^+$) is the total transverse (longitudinal) momentum of the system.

In order to mitigate the computational burden, BLFQ employs an optimized basis for the eigenvalue problem. The BLFQ basis is constructed in terms of the Fock-sector expansion. For each Fock particle (constituent), the basis state is factorized into the longitudinal, transverse and spin (helicity) degrees of freedom. The longitudinal direction ($x^-$) is compactified into a circle of length $2L$, $-L\le x^-\le L$, with periodic (antiperiodic) boundary condition imposed for bosons (fermions) in our application here. Thus, in coordinate space, the longitudinal modes $\psi_k(x^-)$ are given by
\begin{align}
\psi_k(x^-)=\frac{1}{2L}e^{i \frac{\pi}{L}k x^-},
\end{align}
where the wave number $k=\{1,2,3,\cdots\}$ for periodic and $k=\{1/2,3/2,5/2,\cdots\}$ for antiperiodic boundary condition acts as the quantum number for the longitudinal degrees of freedom. 
For periodic boundary condition, we exclude the zero modes ($k=0$).  In the transverse directions, 2D harmonic oscillator (HO) states are adopted as the basis states. In momentum space, we introduce the dimensionless variable $\rho=\mid p^\perp\mid/b$ where $b$ has the dimension of mass. Then the orthonormalized HO wave functions are given by,
\begin{align} 
    \label{eq:2dhowf}
\Phi_{n,m}(\rho,\varphi)=&\sqrt{\frac{2 \pi}{b}}\sqrt{\frac{2n!}{(\mid m\mid+n)!}}~e^{im\varphi}\nonumber\\&\times\rho^{\mid m\mid}e^{-\rho^2/2} L_n^{\mid m\mid}(\rho^2),
\end{align}
with $n$ and $m$ the radial and (2D) angular quantum numbers
and $L_n^{\mid m\mid}(\rho^2)$ are the generalized Laguerre polynomials. 
For the spin degree of freedom, a single quantum number $\lambda$ is used to label the helicity of the particle. In total, each Fock particle is labeled by four quantum numbers, \{$k$, $n$, $m$, $\lambda$\}. The discretized BLFQ basis is orthonormal. 

In order to make numerical calculations feasible, basis truncation is necessary. In BLFQ, basis truncation is performed both at the Fock-sector level and inside each Fock sector. In this work, as an initial application of BLFQ, we make the lowest nontrivial Fock-space truncation for the electron system, namely, we retain only the $\ket{e}$ and $\ket{e\gamma}$ Fock sectors. In each Fock sector, the truncation on transverse degrees of freedom is implemented through the following scheme: we retain only those basis states whose sum of the HO quanta is smaller than a chosen cutoff, $N_\text{max}$, namely,
\begin{align} \sum_i({2n_i+\mid m_i\mid+1}) \le  N_\text{max},
\end{align}
where the sum is over all constituents in that basis state. 

The finiteness of the total HO quanta introduces both transverse ultraviolet (UV) and infrared (IR) cutoffs into the theory~\cite{Coon:2012ab,Furnstahl:2012qg,More:2013rma}. The momentum space HO wave functions fall-off sharply beyond $ p_\perp \propto b\sqrt{N_\text{max}}$ and thus $\Lambda = b\sqrt{N_\text{max}}$ acts as a natural UV cutoff. Taking the Fourier transform of the HO wave functions, one can easily see that the coordinate space wave functions are similar to the momentum space wave functions with the parameter $b$ switching from denominator to the numerator. Thus, the basis states in coordinate space have  maximum support $x_\perp^\text{max}\propto \sqrt{N_\text{max}}/b$. Translated into momentum space, it provides the IR cutoff $\epsilon={1}/{x_\perp^\text{max}}=b/\sqrt{N_\text{max}}$. Therefore, as $N_\text{max}$ increases, for fixed $b$, the UV cutoff increases while the IR cutoff decreases at the same time.

Basis truncation for the longitudinal degrees of freedom is implicit: the imposed (anti-) periodic boundary condition results in discretized longitudinal momenta for the constituents, which leads to a finite number of partitions for a chosen total longitudinal momentum ($P^+$) of the system.


With the BLFQ basis states constructed, we are now in the position to write down the light-front QED Hamiltonian in this basis. Following Ref.~\cite{Zhao:2014xx}, we defer the inclusion of the instantaneous interactions and adopt the following Hamiltonian for this work,
\begin{align}
\label{eq:hami}
    P^- = &\int\!\ud^2x^\LCperp\ud x^\LCm \  \left[\frac{1}{2}\bar{{\Psi}} \gamma^\LCp \frac{m^2+(i\partial^\LCperp)^2}{i\partial^\LCp}\Psi\right.\nonumber\\&\left. + \frac{1}{2} { A}^j (i\partial^\LCperp)^2 { A}^j+ e{j}^\mu {A}_\mu\right],
\end{align}
where $\Psi$ and $A_\mu$ are the fermion and gauge boson field, respectively. $m$ is the bare electron mass and $e$ is the electron charge, which is related to the electromagnetic coupling constant $\alpha=\frac{e^2}{4\pi}$. In order to isolate the eigenstates of the Hamiltonian with lowest transverse center-of-mass motion, we add an appropriate Lagrange multiplier term to the input light-front QED Hamiltonian. This has the effect of shifting the states with excited transverse center-of-mass motion to high mass and the low-lying spectrum comprises states with lowest transverse center-of-mass motion, following the techniques of nuclear many-body theory~\cite{factorization1,factorization2}. The resulting low-lying states can be written as a simple product of internal and center-of-mass motion in the transverse directions.

Upon diagonalizing the QED Hamiltonian in a basis with a chosen total longitudinal momentum ($P^+$) and longitudinal projection of the total angular momentum, the lowest eigenstate is identified as the physical electron, denoted as $\ket{e^\up_\text{phys}}$, where the arrow indicates the helicity of the physical (dressed) electron. The corresponding amplitude can be employed to calculate observables, such as the GPDs. Since in the BLFQ basis we exclude the zero modes for the constituents, we concentrate on the electron GPDs with the (bare) electron longitudinal momentum fraction $x$ only in the region $0<x<1$. In this region and in our truncated basis, the electron GPDs receive contributions from the amplitude in the $\ket{e\gamma}$ sector only. In terms of the BLFQ amplitude of the physical electron state, $\braket{e^\up_\text{phys}}{k_e,n_e,m_e,\lambda_e,k_\gamma,n_\gamma,m_\gamma,\lambda_\gamma}$, we introduce the electron GPDs in the zero-skewness ($\zeta=0$) limit,
\begin{widetext}
\begin{align}
    \label{eq:GPD_H_BLFQ_unrenormalized}
&\phantom{\frac{\Delta^1-i\Delta^2}{2 M}}H\left(x=\frac{k_e}{k_e+k_\gamma},0,t\right)=\sum_{n_e,m_e,\lambda_e,n_\gamma,m_\gamma,\lambda_\gamma} \braket{e'^\up_\text{phys}}{k_e,n_e,m_e,\lambda_e,k_\gamma,n_\gamma,m_\gamma,\lambda_\gamma}\nonumber\\&\qquad\qquad\qquad\qquad\qquad\qquad\qquad\qquad\qquad\qquad\qquad\times\braket{k_e,n_e,m_e,\lambda_e,k_\gamma,n_\gamma,m_\gamma,\lambda_\gamma}{e^\up_\text{phys}};    \\
    \label{eq:GPD_E_BLFQ_unrenormalized}
&\frac{\Delta^1-i\Delta^2}{2 M}E\left(x=\frac{k_e}{k_e+k_\gamma},0,t\right)=\sum_{n_e,m_e,\lambda_e,n_\gamma,m_\gamma,\lambda_\gamma} \braket{e'^\up_\text{phys}}{k_e,n_e,m_e,\lambda_e,k_\gamma,n_\gamma,m_\gamma,\lambda_\gamma}\nonumber\\&\qquad\qquad\qquad\qquad\qquad\qquad\qquad\qquad\qquad\qquad\qquad\times\braket{k_e,n_e,m_e,\lambda_e,k_\gamma,n_\gamma,m_\gamma,\lambda_\gamma}{e^\down_\text{phys}}; 
\end{align}
\end{widetext}
where the summation is over the transverse and helicity quantum numbers of all the constituents. Here $t=\Delta^2=-(\Delta^\perp)^2$ is the squared momentum transfer. Due to the fact that the BLFQ basis adopts discretized longitudinal momenta, we have direct access of the GPDs only at discretized $x=k_e/(k_e+k_\gamma)$. By exploiting the parity symmetry in the transverse plane~\cite{Brodsky:2006ez}, the amplitude for the helicity-down state, $\ket{e^\down_\text{phys}}$, can be inferred from that of the helicity-up state, $\ket{e^\up_\text{phys}}$, as follows,
\begin{widetext}
\begin{align}
   \braket{e^\down_\text{phys}}{k_e,n_e,m_e,\lambda_e,k_\gamma,n_\gamma,m_\gamma,\lambda_\gamma}=(-1)^{m_e+m_\gamma+1}\braket{e^\up_\text{phys}}{k_e,n_e,-m_e,-\lambda_e,k_\gamma,n_\gamma,-m_\gamma,-\lambda_\gamma}.
\end{align}
The amplitude of the final state, $\ket{e'^\up_\text{phys}}$, can also be inferred from that of the initial state, $\ket{e^\up_\text{phys}}$, according to,
\begin{align}
&\braket{e'^\up_\text{phys}}{k_e,n_e,m_e,\lambda_e,k_\gamma,n_\gamma,m_\gamma,\lambda_\gamma}=\sum_{n'_e,m'_e,n'_\gamma,m'_\gamma}\braket{e^\up_\text{phys}}{k_e,n'_e,m'_e,\lambda_e,k_\gamma,n'_\gamma,m'_\gamma,\lambda_\gamma}\nonumber\\&\times\int\!\ud^2k^\LCperp_e \Phi^*_{n'_e,m'_e}(k'^\LCperp_e)\Phi_{n_e,m_e}(k^\LCperp_e)\int\!\ud^2k^\LCperp_\gamma \Phi^*_{n'_\gamma,m'_\gamma}(k'^\LCperp_\gamma)\Phi_{n_\gamma,m_\gamma}(k^\LCperp_\gamma),
\end{align}
\end{widetext}
where for the struck electron $k'^\LCperp_e=k^\LCperp_e-(1- x )\Delta^\LCperp$ and for the spectator photon $k'^\LCperp_\gamma=k^\LCperp_\gamma+(1- x)\Delta^\LCperp$. $\Phi_{n,m}(k^\LCperp)$ is the 2D-HO wave function in momentum space given by Eq.~(\ref{eq:2dhowf}). 

In order for the calculated GPDs to be compared with experimental data,  it is necessary to renormalize the BLFQ results.
Following Ref.~\cite{Zhao:2014xx}, we only perform electron mass renormalization in this work. We adopt a sector-dependent renormalization scheme~\cite{Karmanov:2008br} and adjust the bare electron mass ($m$) in $P^{-}_\text{QED}$ for the matrix elements only in the $\ket{e}$ sector so that the resulting invariant mass of the physical electron state $\ket{e_\text{phys}}$ matches its physical value $M=$0.511\,MeV. The entire process is performed iteratively during the diagonalization of the light-front QED Hamiltonian~(\ref{eq:hami}).

In addition to electron mass renormalization, we need to solve one more issue: the current Fock space truncation violates the condition $Z_1$ = $Z_2$~\cite{Brodsky:2004cx}, which is a consequence of the Ward identity. Here $Z_1$ is the renormalization factor for the vertex coupling the $\ket{e}$ and $\ket{e\gamma}$ sectors which remains unity
in the infinite basis limit with our Fock space truncation.  Now, $Z_2$ is the electron wave-function renormalization which, in light-front dynamics, can be interpreted as the probability of finding a bare electron out of a physical electron:
\begin{align}
    \label{eq:Z_2}
    Z_2 = \sum_{\ket{e}} |\braket{e}{e_\text{phys}}|^2,
\end{align}
where the summation runs over all the basis states in the $\ket{e}$ sector. In our Fock space truncation, $Z_2$ receives a contribution from the quantum fluctuation between the $\ket{e}$ and $\ket{e\gamma}$ sectors and consequently goes to zero in the infinite basis limit.

In order to remedy the resulting artifacts on the observables, we follow the procedure in Ref.~\cite{Zhao:2014xx} and rescale the naive GPDs~(\ref{eq:GPD_H_BLFQ_unrenormalized}-\ref{eq:GPD_E_BLFQ_unrenormalized}) in the region of\footnote{Note that the rescaling for GPDs at $x=1$ needs to be treated separately, since there they receive contributions from the $\ket{e}$ Fock sector.} $0<x<1$ by a factor of $Z^{-1}_2$. We obtain the following rescaled GPDs as our final results,
\begin{align}
    \label{eq:GPDs_BLFQ_renorm}
&\left.H^\text{re}(x,0,t)\right|_{x\in(0,1)}=\left.{Z^{-1}_2}{H(x,0,t)}\right|_{x\in(0,1)},\\
&\left.E^\text{re}(x,0,t)\right|_{x\in(0,1)}=\left.{Z^{-1}_2}{E(x,0,t)}\right|_{x\in(0,1)}.
\end{align}

\section{ Perturbative calculation of the GPDs\label{model}}
 To check the BLFQ results for the GPDs of a dressed electron, we present a perturbative calculation of the electron GPDs using  the  overlap formalism of light-front wave functions.
For this purpose, we use the same Fock space expansion as described in Sec.~\ref{overlap}, see Eq.~(\ref{eq:phys_e_Fock_expansion}).
For zero skewness, the leading contribution (at one loop) comes from the particle number conserving $2\to 2$ process (the single particle sector contributes only for $x=1$ and is a delta function). So, we truncate the Fock space at the two particle sector.  The single particle wave function gives the wave-function renormalization constant and ensures overall probability conservation. The two particle state can again be expanded as \cite{BDH}
\begin{widetext}  
\be
\lefteqn{
\left|\Psi^{\uparrow}_{\rm two \ particle}(P^+,  P_\perp =
0_\perp)\right> =
\int\frac{{\mathrm d} x \, {\mathrm d}^2
           {k}_{\perp} }{\sqrt{x(1-x)}\, 16 \pi^3}
}
\label{vsn1}\\
&&
\left[ \ \ \,
\psi^{\uparrow}_{+\frac{1}{2}\, +1}(x,{k}_{\perp})\,
\left| +\frac{1}{2}\,, +1\, ;\,\, xP^+\, ,\,\, {k}_{\perp}\right>
+\psi^{\uparrow}_{+\frac{1}{2}\, -1}(x,{k}_{\perp})\,
\left| +\frac{1}{2}\, ,-1\, ;\,\, xP^+\, ,\,\, {k}_{\perp}\right>
\right.
\nonumber\\
&&\left. {}
+\psi^{\uparrow}_{-\frac{1}{2}\, +1} (x,{k}_{\perp})\,
\left| -\frac{1}{2}\,, +1\, ;\,\, xP^+\, ,\,\, {k}_{\perp}\right>
+\psi^{\uparrow}_{-\frac{1}{2}\, -1} (x,{k}_{\perp})\,
\left| -\frac{1}{2}\, ,-1\, ;\,\, xP^+\, ,\,\, {k}_{\perp}\right>\
\right] \ ,
\nonumber
\ee
where the two numbers in the subscript of the wave functions denote the helicities of the
bare electron and the photon ($\lambda_e$ and $\lambda_{\gamma}$), respectively. $x$ is the longitudinal momentum fraction for the bare electron and 
 ${k}_{\perp}$ refers to its relative momentum.  The 
 two-particle states $\mid \lambda_ e, \lambda_ \gamma;  p^+, {k}_{\perp} \rangle$ are normalized as
\be
\langle {\lambda'}_{ e}, \lambda'_{ \gamma};  p'^+, k'_\perp  \mid  \lambda_e, \lambda_\gamma; p^+, k_\perp \rangle=16\pi^3 p^+\delta(p'^+-p^+)\delta^2(k'_\perp-k_\perp)\delta_{\lambda'_e,\lambda_e} \delta_{\lambda'_\gamma,\lambda_\gamma}.
\ee
\end{widetext} 
Similarly, one can write down the expansion for the helicity-down physical electron state.
 
In BLFQ, the two particle wave functions are evaluated nonperturbatively and numerically and then are used to calculate the GPDs. To verify  consistency, we have used the same Fock space truncation in both perturbative and BLFQ calculations.
  
For the perturbative calculation, we use the wave functions~\cite{BDH}:
\be
\left
\{ \begin{array}{l}
\psi^{\uparrow}_{+\frac{1}{2}\, +1} (x,{k}_{\perp})=-{\sqrt{2}}
\ \frac{-k^1+{i} k^2}{x(1-x)}\,
\varphi(x,k_\perp) \ ,\\
\psi^{\uparrow}_{+\frac{1}{2}\, -1} (x,{k}_{\perp})=-{\sqrt{2}}
\ \frac{k^1+{i} k^2}{1-x }\,
\varphi(x,k_\perp) \ ,\\
\psi^{\uparrow}_{-\frac{1}{2}\, +1} (x,{k}_{\perp})=-{\sqrt{2}}
\ (M-{m\over x})\,
\varphi(x,k_\perp) \ ,\\
\psi^{\uparrow}_{-\frac{1}{2}\, -1} (x,{k}_{\perp})=0\ ,
\end{array}
\right.
\label{vsn2}
\ee
\be
\varphi (x,{k}_{\perp}) = \frac{e}{\sqrt{1-x}}\
\frac{1}{M^2-{{k}_{\perp}^2+m^2 \over x}
-{{k}_{\perp}^2+m_\gamma^2 \over 1-x}}\ .
\label{wfdenom}
\ee
where $M$ is the physical electron mass, $m$ is the bare electron mass, and  $m_\gamma$ is the photon mass.  In perturbative QED, $M=m$ and  we keep a small nonzero photon mass which acts as an IR cutoff.  Though we don't have any IR divergence in the GPDs, the purpose of the nonzero photon mass is to compare with  the intrinsic IR regulator in the HO basis used in BLFQ, see Sec.\ref{blfq}.

The GPDs  of the electron in  perturbation theory are 

\be
H(x,0,t)&=&\frac{\alpha}{2 \pi}\Big[\frac{1+x^2}{1-x}\ln\vert\frac{\Lambda^2}{A}\vert \nonumber\\
&& + \frac{1+x^2}{1-x}A I_1+M^2(1-x)^3 I_1\nonumber\\&&-\frac{1}{2}(1-x)(1+x^2)\Delta_\perp^2 I_1\Big],\\
E(x,0,t)&=&\frac{\alpha}{\pi} M^2 x(1-x)^2 I_1.
\ee
where $ A= M^2 x(1-x)-m^2(1-x)- m_\gamma^2 x = -M^2(1-x)^2-m_\gamma^2 x ~({\rm as}~ M=m)$, $\Lambda$ is the UV cutoff, and  for $\zeta=0$, the square of momentum transferred $t=\Delta^2=-\Delta_\perp^2$. 
The integration $I_1$ is  defined as
\be
I_1=\int_0^1dy \frac{\Lambda^2}{\beta(x,y)(\beta(x,y)+\Lambda^2)},
\ee
where $\beta(x,y)= y(1-y)(1-x)^2\Delta_\perp^2+M^2(1-x)^2+m_\gamma^2 x$.
Note that the integration $I_1$ is finite in the limit $\Lambda\to\infty$, but we show the explicit UV cutoff dependence to compare  with the  BLFQ results.

\begin{figure}[t]
\includegraphics[width=0.45\textwidth]{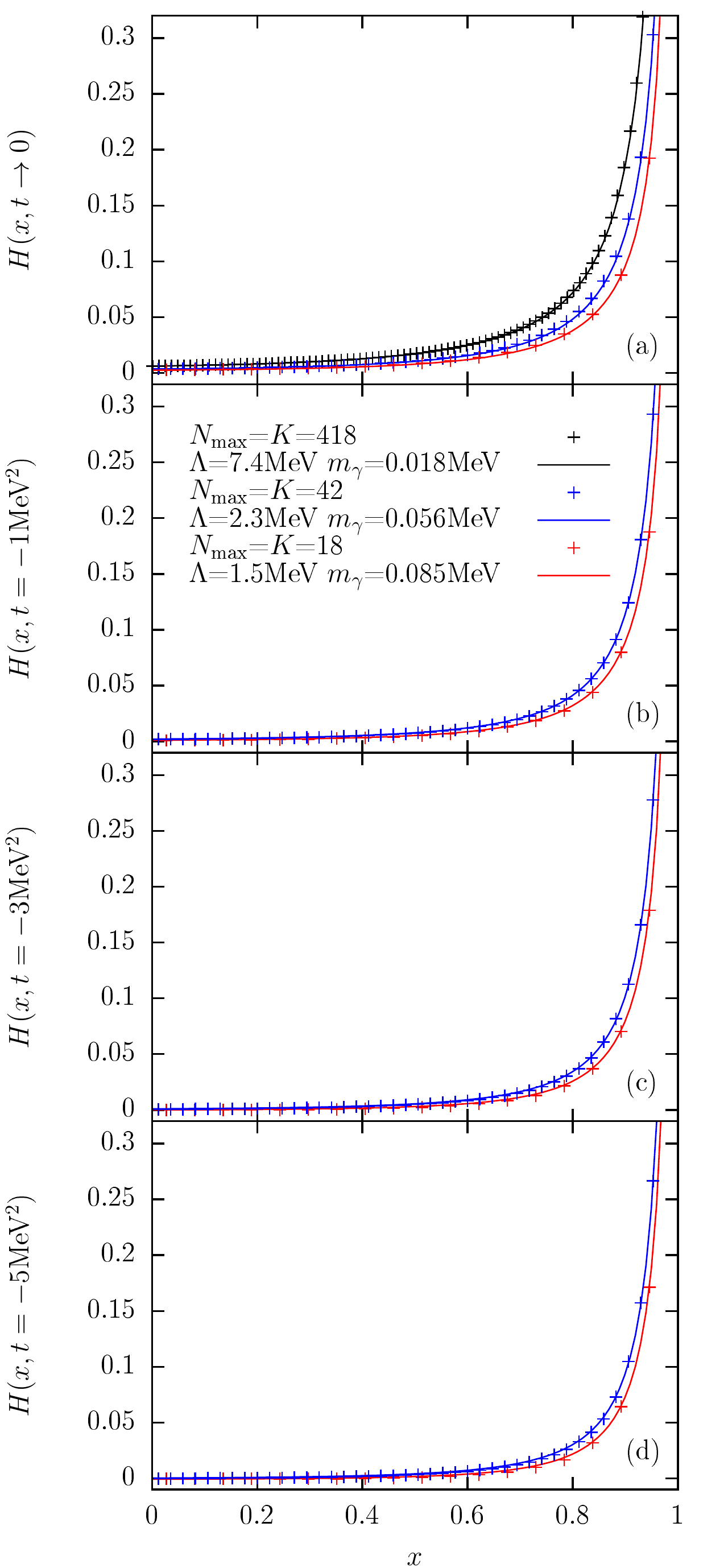}
\caption{\label{fig1} (Color online)  GPD  $H(x,t)$  for different values of $t$: $t\to 0$ (a), $t=-1$MeV$^2$ (b), $t=-3$MeV$^2$ (c),
  $t=-5$MeV$^2$ (d). For $t\to 0$, $H(x,t\to 0)$ is the  PDF for the dressed electron. Solid lines represent the perturbative results; data points are the BLFQ results.} 
\end{figure}
\section{Results\label{results}}
\begin{figure}[t]
\includegraphics[width=0.45\textwidth]{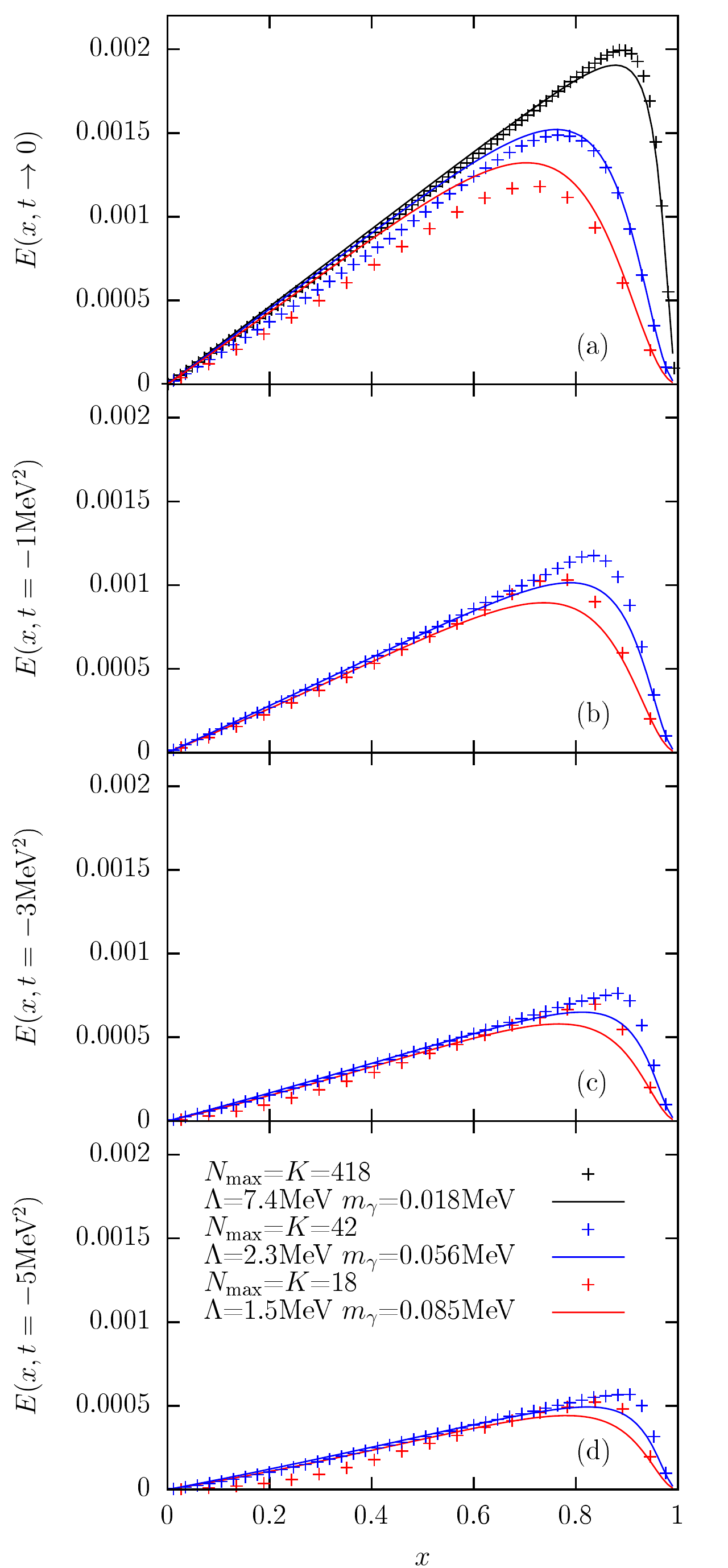}
\caption{\label{fig2} (Color online)  GPD $E(x,t)$ for different values of $t$: $t\to 0$ (a), $t=-1$MeV$^2$ (b), $t=-3$MeV$^2$ (c), $t=-5$MeV$^2$ (d). Solid lines represent the perturbative results; data points are the BLFQ results.} 
\end{figure}
Here we present the numerical results from  both non-perturbative (BLFQ) and perturbative methods calculated with $\alpha=1/137.036$ and $M=0.511$\,MeV. For the perturbative model, 
the photon mass $m_\gamma$ and the UV cutoff $\Lambda$ are tuned to match with the IR cutoff $\epsilon=b/\sqrt{N_\text{max}}$ and the UV cutoff $b\sqrt{N_\text{max}}$ in BLFQ, \ie, we set $m_\gamma=\epsilon=b/\sqrt{N_\text{max}}$ and $\Lambda=b\sqrt{N_\text{max}}$. 

In Fig.~\ref{fig1}, we have presented the comparison of GPD $H(x,t)=H(x,0,t)$ (since we are considering here only $\zeta=0$, we suppress it in the arguments of  all GPDs) for  different values of total momentum transferred $t$. Note that for $\zeta=0$, $t=\Delta^2=-\Delta_\perp^2$ is a negative quantity. 
The GPD $H(x,t)$ reduces to the ordinary parton distribution function (PDF) in the forward limit $t\to 0$.
The dependence on the photon mass or IR cutoff in the PDF or GPD $H(x,t)$ is not very significant.  
In Fig.~\ref{fig1},  $H(x,t\to 0)$  presents the PDF for the physical (dressed) electron. The BLFQ result agrees well with the perturbative calculation which is expected, since
in the Fock-sector truncation allowing only for the quantum fluctuation into the $\ket{e\gamma}$ sector, the resulting nonperturbative light-front wave function encodes the identical information on the structure of the physical electron, compared to that from leading-order perturbation theory. The higher-order contributions only contribute to the electron wave-function renormalization factor, $Z_2$.  However, due to the fact that we use a truncated 2D HO basis for the BLFQ calculations, and a plane wave basis for the perturbative calculations, we do anticipate (small) differences between the BLFQ and perturbative results.

Although it appears that $H(x,t)$ diverges at $x=1$, we should remember that there is a single particle ($\ket{e}$) contribution to the GPD at exactly $x=1$ which, with proper normalization, cancels the divergence
and produces the desired normalization of the GPD $\int_0^1H(x,0)dx=F_1(0)=1$ for the electron \cite{CM}.
 
In Fig.~\ref{fig2} we compare the BLFQ results for GPD $E(x,t)$  with the perturbative results for different values of $t$. Note that there is no divergence at $x=1$ in $E(x,t)$. In the perturbative calculation the photon mass $m_\gamma$ can safely be set to zero and the UV cutoff $\Lambda$ can be taken to the infinite limit. However, in the BLFQ, the IR and UV cutoffs are intrinsic to the formalism as $b\ne0$ and $N_\text{max}$ is finite. We therefore retain the nonzero photon mass $m_\gamma=b/\sqrt{N_\text{max}}$ and also set $\Lambda=b\sqrt{N_\text{max}}$ in the perturbative calculation for consistency with BLFQ. Overall, the agreement between the BLFQ and perturbative results seems reasonable and it improves as the BLFQ basis size ($N_\text{max}$ and $K$) increases. Minor mismatches are visible in the region of $x$ approaching one. This is because unlike GPD $H(x,t)$, GPD $E(x,t)$ is very sensitive to the IR cutoff, especially at around $x\sim 1$. Although in this calculation the IR cutoffs in BLFQ and in perturbation theory are ``matched'', they are, however, not sharp cutoffs in either method and have their distinct cutoff profiles.

\begin{figure}[t!]
\centering
\includegraphics[width=0.48\textwidth]{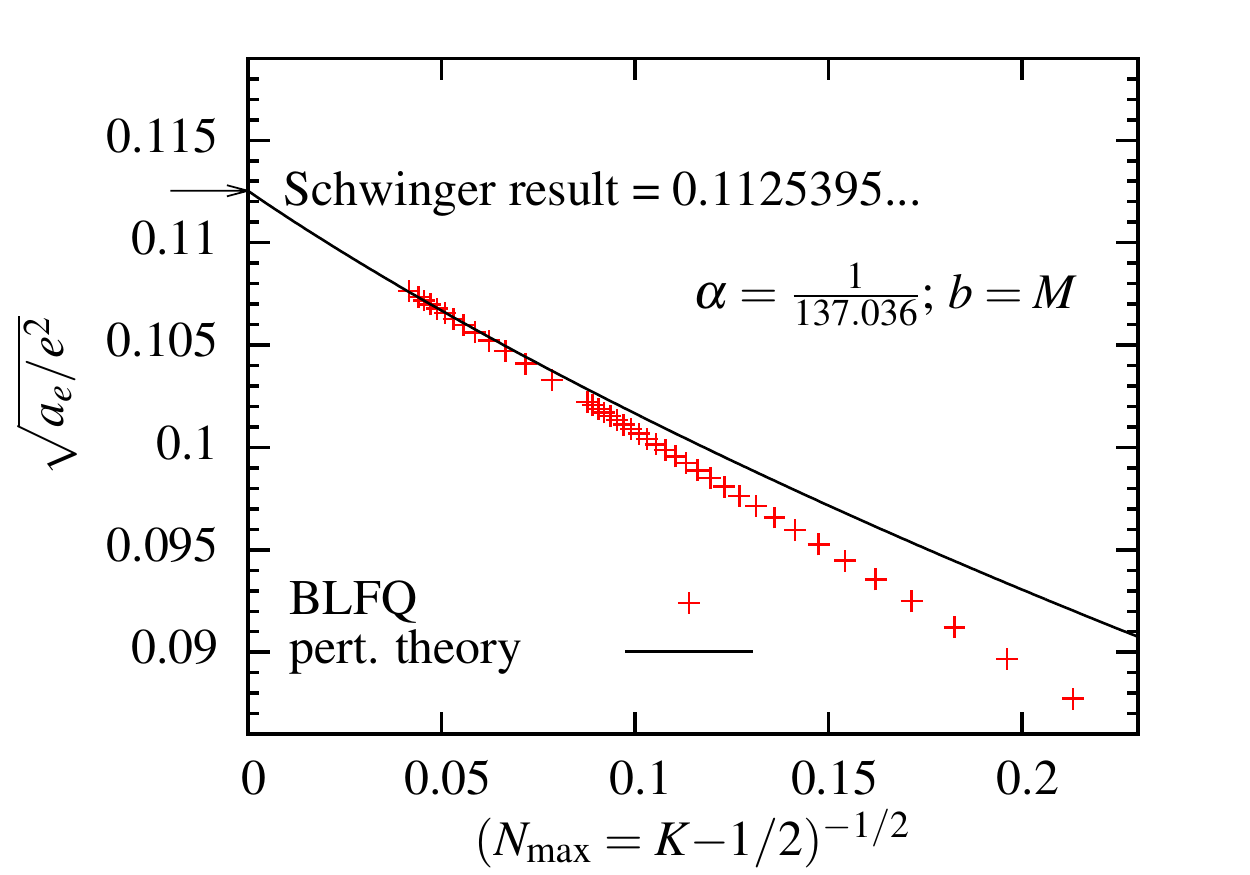}
\caption{\label{fig3} (Color online) 
Comparison between the electron anomalous magnetic moment, $a_e=\frac{g-2}{2}$, evaluated in perturbation theory and that in BLFQ. 
The $x$-axis indicates (the inverse square root of) the basis truncation parameter $N_\text{max}{=}K{-}1/2$ and the $y$-axis is the (square root of) electron anomalous magnetic moment $a_e$ (normalized to electron charge $e$). Solid lines represent the perturbative results; data points are the BLFQ results. The largest BLFQ calculation is at $N_\text{max}=K-1/2=$578 with the basis dimensionality of 18,650,072,547.} 
\end{figure}
In Fig.~\ref{fig3} we have shown a comparison of the BLFQ and perturbative results for electron anomalous magnetic moment,  $a_e=(g-2)/2$, 
which is the integral of the GPD $E(x,t\to 0)$ over $0<x<1$. Here $g$ is the gyromagnetic ratio for the electron. Again, the IR and UV cutoffs are matched between the BLFQ and perturbative results. As $N_\text{max}=K{-}1/2$ increases, the agreement between the perturbative and BLFQ results improves. In the limit of $N_\text{max}=K{-}1/2\to\infty$, the BLFQ result nicely extrapolates to the Schwinger value, see Ref.~\cite{Zhao:2014xx} for a more detailed description on the electron anomalous magnetic moment calculation in BLFQ.

\section{Summary and Conclusions}
Basis Light-front Quantization has recently been developed as a promising non-perturbative light-front technique.
In this paper, we have investigated the GPDs in the BLFQ method.   For this purpose, we have considered a physical (dressed) electron in QED  so that  our nonperturbative results can be compared with perturbative calculations.   The physical  electron state has been expanded into its Fock space basis to evaluate 
the GPDs $H(x,t)$ and $E(x,t)$ for zero skewness which, in leading twist, are given by the particle number conserving processes.
Our initial study in the lowest nontrivial  Fock sectors shows that with a proper renormalization procedure and a rescaling of the naive GPDs correcting the artifacts introduced by the Fock space truncation, the BLFQ results are consistent with the perturbative results.  The main goal of this study in the BLFQ approach is to establish the foundation for investigating the  GPDs for nucleons which are highly nonperturbative.
To investigate the strong coupling physics, we need to include higher Fock sectors   as well as to increase the total quanta ($N_\text{max}$ and $K$) for the BLFQ basis states.
The HO basis employed in BLFQ works well for systems with  localized wave functions (bound states). Since QCD has confinement, we expect that the convergence in the HO basis will be better compared to QED.

{\bf Acknowledgements:}
We thank Asmita Mukherjee, Matthias Burkardt and Stanley J. Brodsky for fruitful discussions.
This work  was supported in part by the DOE Grant
Nos. DE-FG02-87ER40371,
DESC0008485 (SciDAC-3/NUCLEI), and by US NSF grant 0904782. H.~H is supported by US DOE Contract Number DE-FG02-93ER40771. A portion of the computational resources were provided by the National Energy Research Scientific
Computing Center (NERSC), which is supported by the Office of Science of the U.S. Department of Energy under Contract
No. DE-AC02-05CH11231. 

\end{document}